\documentclass[runningheads]{llncs}
\usepackage[T1]{fontenc}
\usepackage{graphicx}
\usepackage{graphicx}
\usepackage{amsmath}
\usepackage{amssymb}
\usepackage{booktabs}
\usepackage{pifont}
\usepackage{colortbl}
\usepackage{algorithm}
\usepackage{algorithmic}
\usepackage{bbding}
\usepackage{soul}
\usepackage{bibunits}
\usepackage{bm}
\usepackage{array}
\usepackage[dvipsnames]{xcolor} 
\usepackage{pdflscape}
\usepackage{makecell}
\usepackage{framed,multirow}
\usepackage{threeparttable}
\usepackage{amssymb}
\usepackage{pifont}
\usepackage{algorithm}
\usepackage{listings}
\usepackage{pythonhighlight}
\usepackage{float}
\usepackage{mathtools}
\usepackage{cuted}
\DeclareUnicodeCharacter{2212}{-}
\makeatletter
\newcommand\footnoteref[1]{\protected@xdef\@thefnmark{\ref{#1}}\@footnotemark}
\makeatother

\newcolumntype{P}[1]{>{\centering\arraybackslash}p{#1}}
\newlength\savewidth

\newcommand{\etal}{\mbox{et al.}}

\def\arrvline{\hfil\kern\arraycolsep\vline\kern-\arraycolsep\hfilneg}
\definecolor{mygray}{gray}{.9}
\definecolor{Highlight}{HTML}{39b54a}  

\newcolumntype{x}[1]{>{\centering\arraybackslash}p{#1pt}}
\newcolumntype{z}[1]{>{\raggedright\arraybackslash}p{#1pt}}

\definecolor{citecolor}{HTML}{0071BC}
\definecolor{linkcolor}{HTML}{ED1C24}
\usepackage[pagebackref,breaklinks,colorlinks,citecolor=citecolor]{hyperref}

\usepackage[T1]{fontenc}

\newcommand{\ourmodel}{{\fontfamily{ppl}\selectfont Pixel2Cancer}}
\definecolor{iblue}{rgb}{0.06, 0.75, 1.0}
\definecolor{ired}{rgb}{0.8588, 0.2666, 0.2156}

\begin{document}
%

\title{From Pixel to Cancer:\\Cellular Automata in Computed Tomography}

\titlerunning{From Pixel to Cancer}
\author{Yuxiang Lai\inst{1,2} \and 
Xiaoxi Chen\inst{3} \and
Angtian Wang\inst{1} \and \\
Alan Yuille\inst{1} \and
Zongwei Zhou\thanks{Correspondence to: Zongwei Zhou (\href{mailto:zzhou82@jh.edu}{\texttt{zzhou82@jh.edu}})}\inst{1}}
\authorrunning{Y. Lai et al.}
\institute{John Hopkins University \and
Southeast University \and
University of Illinois Urbana-Champaign 
}
\maketitle              
\begin{abstract}
AI for cancer detection encounters the bottleneck of data scarcity, annotation difficulty, and low prevalence of early tumors. Tumor synthesis seeks to create artificial tumors in medical images, which can greatly diversify the data and annotations for AI training. However, current tumor synthesis approaches are not applicable across different organs due to their need for specific expertise and design. This paper establishes a set of generic rules to simulate tumor development. Each cell (pixel) is initially assigned a state between zero and ten to represent the tumor population, and a tumor can be developed based on three rules to describe the process of growth, invasion, and death. We apply these three generic rules to simulate tumor development---from pixel to cancer---using \textit{cellular automata}. We then integrate the tumor state into the original computed tomography (CT) images to generate synthetic tumors across different organs. This tumor synthesis approach allows for sampling tumors at multiple stages and analyzing tumor-organ interaction. Clinically, a reader study involving three expert radiologists reveals that the synthetic tumors and their developing trajectories are convincingly realistic. Technically, we analyze and simulate tumor development at various stages using 9,262 raw, unlabeled CT images sourced from 68 hospitals worldwide. The performance in segmenting tumors in the liver, pancreas, and kidneys exceeds prevailing literature benchmarks, underlining the immense potential of tumor synthesis, especially for earlier cancer detection. The code and models are available at \href{https://github.com/MrGiovanni/Pixel2Cancer}{\texttt{https://github.com/MrGiovanni/Pixel2Cancer}}

\end{abstract}

\setcounter{footnote}{0} 

\section{Introduction}\label{sec:introduction}

Training AI models for image segmentation needs extensive datasets with detailed per-pixel annotations To reduce manual annotation efforts, \textit{data synthesis} is a compelling approach for generating enormous synthetic data and annotations for both training~\cite{lyu2022pseudo,hu2023label} and evaluating~\cite{hu2023synthetic} AI models. This is particularly attractive for segmenting small tumors, which are clinically important to detect at the early stage but rarely occur in public datasets because they are difficult for radiologists to identify.

Preliminary studies have been successful in synthesizing specific disease conditions in medical images, such as synthesizing pulmonary inflammatory lesions in COVID-19~\cite{lyu2022pseudo}, lung nodules~\cite{han2019synthesizing}, abdominal tumors~\cite{jin2021free,chen2024towards}, and brain tumors~\cite{wyatt2022anoddpm}. However, key challenges remain: (i) generating highly realistic tumors, (ii) synthesizing tumors that generalize across multiple organs, and (iii) leveraging synthetic tumors for AI model training. Moreover, while some mathematical studies~\cite{forster2017development,menze2011generative} have modeled tumor development in simplified hypothetical environments, they cannot generate realistic synthetic data for model training.

In this paper, we revisit a classic technique, cellular automata~\cite{wolfram1983statistical}, incorporating novel designs to generate realistic tumors of various shapes. In cellular automata, the basic element is the \textit{cell}, which our work refers to as a single \textit{pixel} in the computed tomography (CT) image. We propose \ourmodel, a framework that can generate synthetic tumors guided by three generic rules we have developed\footnote{We develop these generic rules by imitating tumor behavior based on medical knowledge (detailed in §\ref{sec:pixel2cancer}), including tumor growth and death~\cite{forster2017development}, tissue invasion~\cite{frieboes2010three}, and interactions with the tumor microenvironments~\cite{subramanian2019simulation}.}. As a result, \ourmodel\ can generate multiple stages of tumor development (illustrated in~\figureautorefname~\ref{fig_pipeline}), simulate the interaction between tumors and their surrounding environments, exhibit applicability across various organs, and potentially contribute to the estimation of tumor prognosis in longitudinal data.

We have evaluated the \ourmodel\ from two perspectives. \textit{Clinically}, reader studies involving three expert radiologists have been conducted, and results have been so convincing that even medical professionals with over ten years of experience can mistake them for real tumors (\tableautorefname~\ref{tab:reader_studies}). \textit{Technically}, we have generated synthetic tumors on healthy data to train the segmentation model. The performance in segmenting tumors in the liver, pancreas, and kidneys exceeds previous benchmarks and even model training on real data (Appendix \tableautorefname~\ref{tab:comparation_real}), underlining the immense potential of tumor synthesis, especially for earlier cancer detection/diagnosis (Appendix~\figureautorefname~\ref{fig:early_tumors}). 

This performance is attributable to our key observation: Many tumors, regardless of origin, share underlying growth, invasion, and interaction dynamics. Our generic rules effectively capture these shared processes, enabling tumor synthesis across diverse organs. Benign tumors and early-stage cancers usually grow in a contained manner, pushing against surrounding tissues rather than invading them. In contrast, advanced cancers often exhibit invasive growth, infiltrating surrounding tissues and potentially spreading to distant parts of the body (metastasis)\footnote{Using liver cancer as an example, early-stage liver cancer commonly presents as small nodules with a relatively regular shape. In contrast, advanced liver cancer frequently demonstrates infiltration of intrahepatic blood vessels and bile ducts, or distant metastasis. Similar patterns are observed in other solid tumors as well~\cite{m2021use}.}. Furthermore, we found that the 9,262 raw CT images we used were captured during the venous phase, during which a significant proportion of tumors appeared hypo-intense (dark). Therefore, we chose to simulate this typical presentation, closely aligning with clinical practice. This observation implies the existence of fundamental tumor development principles that transcend organ-specific variations. We hereby contrast \ourmodel\ with \textbf{related work} to show our innovations and novelties.

\begin{enumerate}
    \item \textbf{Requiring no manual annotation.} Learning-based approaches, such as GAN~\cite{goodfellow2014generative} and Diffusion~\cite{ho2020denoising}, excel in learning tumor representations but require abundant paired tumor data for effective CT image generation. Moreover, the generation process requires extra manual efforts, including the creation of masks to indicate tumor locations and shapes~\cite{jin2021free}. 

    \item \textbf{Simulating tumors development.}
    None of the existing synthetic approaches can simulate tumor development, especially in the intricate processes of proliferation and invasion~\cite{harpold2007evolution}. The complexity of these processes, influenced by the surrounding environment~\cite{tanase2015complexity}, makes it difficult to generate synthetic tumors when interacting with organ tissues and structures, particularly when generalizing to different organs.

    \item \textbf{Synthesizing tumors across organs.} Modeling-based approaches leverage specialized design and domain expertise to simulate tumor appearances. Although the utilization of modeling-based synthetic tumors eliminates the need for manual annotation~\cite{hu2023label}, they require significant effort for designing proper tumor characteristics customized to a specific organ, therefore limiting generalization, especially across different organs.
\end{enumerate}

\section{\ourmodel}\label{sec:pixel2cancer}

\begin{figure*}[t]
	\centering
\includegraphics[width=\linewidth]{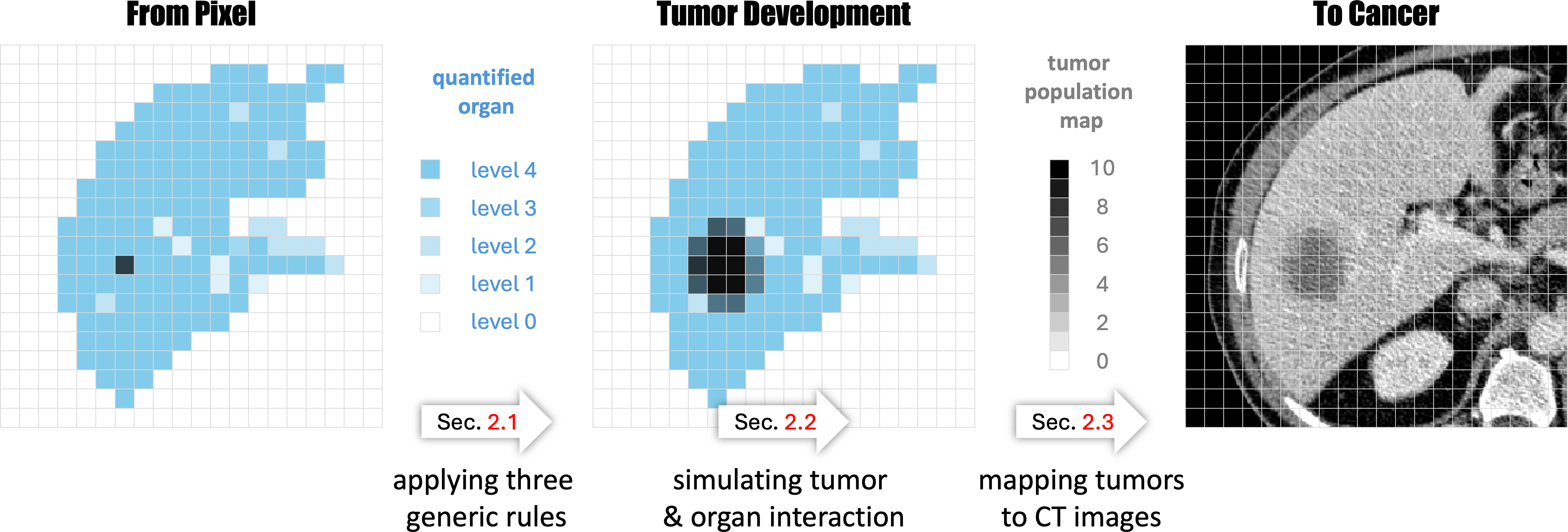}
	\caption{Initially, we quantify the organ from CT intensity to create the quantified organ and select a pixel as the starting point. Subsequently, we apply three generic rules to simulate tumor development based on the quantified organ, including growth, interaction, and death (\S\ref{sec:three_rules}). Then, using these generic rules, \ourmodel\ simulates tumor development and records simulation results using a tumor population map (\S\ref{sec:tumor_organ_interaction}). Finally, we generate tumors in CT scans through the mapping function according to the tumor population map and CT intensity (\S\ref{sec:map_back_ct}).
 }  \label{fig_pipeline}
\end{figure*}

\textbf{Overview.} We begin by quantifying an organ (e.g., liver) based on CT values, assigning each pixel to four levels: level 1 (softest) to level 4 (hardest), with vessels/boundaries as level 0. Next, a tumor is generated from an arbitrary pixel of the organ tissue (levels $>$ 0). Inspired by cellular automata, we design three generic rules in \S\ref{sec:three_rules}: (1) growth, (2) invasion, and (3) death. These rules oversee the tumor development and interaction with the quantified organ (see \S\ref{sec:tumor_organ_interaction}), maintaining a dynamically updated tumor population map. Finally, we translate this map back to the original CT image (see \S\ref{sec:map_back_ct}), modifying the CT values based on the original values and the tumor population---a greater tumor population denotes a smaller CT value (visually darker). As shown in \figureautorefname~\ref{fig_pipeline}, our \ourmodel\ enables the generation of tumors at various stages in CT images.

\begin{figure*}[t]
	\centering
    \includegraphics[width=\linewidth]{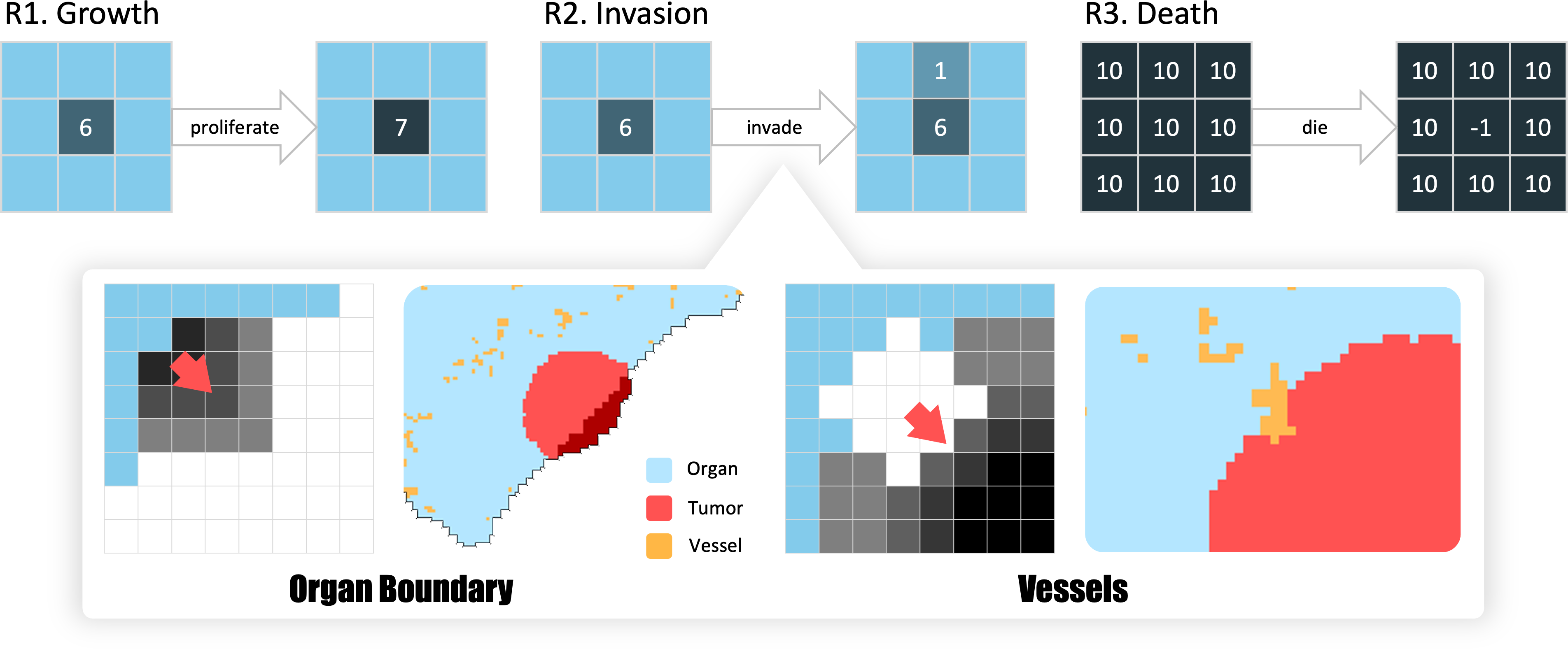}
	\caption{\textbf{R1. Growth:} Tumor cells can proliferate themselves ($\text{self-state} +1$) with probability. \textbf{R2. Invasion:} Tumor cells can invade neighboring cells ($\text{neighbor-state} +1$). In this process, we simulate interactions among tumors, organ tissues, vessels, and boundaries. At the bottom line, we present cases where tumors are compressed by organ boundaries and vessels. \textbf{R3. Death:} Tumor cells surrounded by a full population of neighboring cells ($\text{state}=10$) will undergo cell death ($\text{self-state}\leftarrow -1$).}
 \label{fig_growth}
\end{figure*}

\subsection{Designing Three Rules for Tumor Development}\label{sec:three_rules}

Since CT pixel scales are larger than cell scales, we introduce a \textit{tumor population map} 
to quantify the tumor population state within each cell (\textit{pixel}). The state ranges from 0 (no tumor) to 10 (maximal tumor population), updating over time based on the states of neighboring cells. A key challenge in cellular automata is designing the rules overseeing state evolution. To initiate tumor development, we randomly select a cell in the organ region and assign its $\text{state}=1$. We then defined three generic rules to simulate tumor shape, texture, and behaviors:

\begin{itemize}
    
    \item \textbf{R1: Growth.} Cells with a state greater than 0 but less than 10 can accumulate with a probability ($\text{self-state} +1$). This rule simulates the proliferative behavior of the tumor.
    
    \item \textbf{R2: Invasion.} Cells with a state greater than 0 can incentivize one of its neighbors to grow ($\text{neighbor-state} +1$). This rule simulates the invasive behavior of the tumor.
    
    \item \textbf{R3: Death.} Cells with a state of 10 surrounded by neighbors with 10 can die with a probability ($\text{self-state}\leftarrow -1$). This rule simulates the effect of crowding in the tumor, which leads to the death of tumor cells.

\end{itemize}

R1 and R3 are straightforward, but R2's complexity arises from its potential to influence neighboring cells. Tumor invasion depends not only on its population but also on the nature of neighboring cells (e.g., tissue, vessels, boundaries, etc.). To address this, the next section will describe how we model the interaction between organs and tumors during the invasion process (R2).

\subsection{Simulating Interaction Between Organs and Tumors}\label{sec:tumor_organ_interaction}


\smallskip\noindent\textbf{Quantified Organ.} 
CT images typically exhibit a wide range of intensities, typically ranging from $-1000$ to $1000$ Hounsfield Units (HU), which poses challenges for simulating interactions between synthetic tumors and surrounding tissues. To simplify the simulation process, we convert the organ in CT images into four levels, resulting in the quantified organ. As shown in~\figureautorefname~\ref{fig_pipeline}, the quantified map categorizes the CT intensity into four levels, spanning from low to high intensity (level 1 to 4), with vessels/boundaries assigned to level 0. Subsequently, \ourmodel\ utilizes this quantified organ as the simulation environment to simulate growth and interaction (\figureautorefname~\ref{fig_growth}). 


\smallskip\noindent\textbf{Growth and Interaction:}
As mentioned in \S\ref{sec:three_rules}, the tumor will proliferate itself on the tumor population map (self-state $+1$). Increasing populations lead to greater space occupation, resulting in interactions with and invasion of surrounding tissues. Crowded tumor cells will randomly choose directions and attempt to invade.
When interacting with organs, tumors can directly invade normal tissues (level 1,2,3) at different rates. However, the presence of physical barriers such as vessels, boundaries (level 0), and high-density organ tissues (level 4) can impede tumor growth. In this case, crowded tumor cells (state = 10) exert pressure as they attempt to invade surrounding tissues. The pressure levels range from 0 to n, indicating the number of invasion attempts (\figureautorefname~\ref{fig_growth}). When the pressure of the target direction reaches the threshold, tumor cells can invade these high-density tissues.
In~\figureautorefname~\ref{fig_growth}, we also simulate tumor cell death, influenced by factors such as the vascular feeding environment and cell size~\cite{rieger2015integrative}. Our model corresponds to cells surrounded by maximal population neighbors (neighbor-states = 10), and certain evolutionary steps (of sufficient size) can lead to cell death.

\subsection{Mapping Tumors to CT Images}\label{sec:map_back_ct}

After simulating tumor development, the resulting tumor population map needs to be mapped back into CT images (\figureautorefname~\ref{fig_pipeline}). The mapping process is based on a mapping function, eliminating the need for annotations and training processes. It integrates information from the tumor population map and original CT intensity. Tumor population maps guide the mapping process to modify CT intensity values, reflecting different tumor states, including dead cells ($\text{state}=-1$), tumor cells ($\text{state}=10$), normal cells, and blurry boundaries where two cell types mix ($0<\text{state}<10$).
To identify the intensity of each tumor pixel, we incorporate typical intensity differences between organ tissues and tumors. Then, we can calculate the intensity by considering the proportion of tumor populations within pixels.
To create realistic tumor textures, we introduce 3D Gaussian noise with predefined mean intensity and standard deviation as organ parenchyma.

\begin{table}[t]
    \centering
    \scriptsize
    \caption{
    \textbf{Visual Turing Test} involved three experts each evaluating 150 CT images, with 50 per organ. They were tasked with categorizing each CT image as either \textit{real}, or \textit{synthetic}. A lower specificity score indicates a higher number of synthetic tumors being identified as real. We also provide typical examples of synthetic tumors recognized correctly and incorrectly. The red dot indicates the human expert believes this is a synthetic tumor, while the green dot indicates they believe it is a real tumor.
    }
    \begin{tabular}{p{0.18\linewidth}p{0.23\linewidth}P{0.18\linewidth}P{0.18\linewidth}P{0.18\linewidth}}
    \toprule
     & metric& liver & pancreas & kidneys \\
    \midrule
    \multirow{3}{*}{\makecell[l]{\textbf{R1}\\\textit{3-year}\\\textit{experience}}} & sensitivity (\%) &100 &95.0 & 95.5\\
     & specificity (\%) &27.3 & 22.7 & 26.7\\
     & accuracy (\%)  & 60.9 & 57.1& 67.6\\
    \midrule
    \multirow{3}{*}{\makecell[l]{\textbf{R2}\\\textit{7-year}\\\textit{experience}}} & sensitivity (\%) &94.7 & 87.5 & 90.0\\
     & specificity (\%) &47.8 & 47.4 &56.3\\
     & accuracy (\%) &69.1 & 65.7 & 75.0\\
    \midrule
    \multirow{3}{*}{\makecell[l]{\textbf{R3}\\\textit{10-year}\\\textit{experience}}} & sensitivity (\%) & 100 & 100 & 100\\
     & specificity (\%)  & 45.4 & 55.6 & 57.9\\
     & accuracy (\%)  & 68.4 & 72.4 & 75.8\\
    \bottomrule
    \end{tabular}
    
    \begin{tablenotes}
        \item positives: real tumors ($N$ = 25); negatives: synthetic tumors ($N$ = 25). 
    \end{tablenotes}
    \vspace{2mm}
    \begin{tabular}{c}
        \includegraphics[width=\linewidth]{fig_reader_study.png}
    \end{tabular}
    
    \label{tab:reader_studies}
\end{table}

\section{Fake Tumors, Real Results}\label{sec:result}
\subsection{Experiment}
\label{sec:dataset}
\noindent\textbf{Datasets:} 
For \textit{training}, we synthesize tumors on healthy CT data. The healthy liver dataset comprises images with healthy livers assembled from CHAOS~\cite{kavur2021chaos}, BTCV~\cite{landman2015}, Pancreas-CT~\cite{TCIA_data}, and healthy subjects in LiTS. The healthy pancreas dataset comprises images from Pancreas-CT and BTCV. The healthy kidney dataset consists of images from BTCV, WORD~\cite{luo2021word}, and Abdomenct-1k~\cite{ma2021abdomenct}. For \textit{evaluation}, Liver tumor annotations are from LiTS~\cite{bilic2019liver}. Pancreas tumors are annotated in MSD (Task07 Pancreas)~\cite{simpson2019large}. Kidney annotations are from KiTS~\cite{heller2019kits19}. We use 5-fold cross-validation to evaluate the performance. 

\smallskip\noindent\textbf{Implementation:}
\textit{Cellular Automata:} We conduct simulation by implementing 3D Cellular Automata package using CUDA. In the 3D space, the CA kernel size is $(3,3,3)$, with cell growth frequency set to $1$ in each iteration. We employed $1024$ blocks with $64$ threads for multiprocessing. Compared to standard CPU implementations, our CUDA-based approach achieves a speedup of $500\times$. \textit{Tumor segmentation:} Our segmentation code is based on the \href{https://monai.io}{MONAI} framework using U-Net~\cite{ronneberger2015u}, Swin-UNETR~\cite{hatamizadeh2022swin}, and nnU-Net~\cite{isensee2021nnu} as backbones. During training, random patches of size $96 \times 96 \times 96$ are cropped from 3D CT images. Models are trained for $2,000$ epochs with a base learning rate of $0.0002$, using a batch size of $2$ per GPU and employing linear warm-up and cosine annealing learning rate schedule. For inference, we use the sliding window strategy with $0.75$ overlapping.

\smallskip\noindent\textbf{Hyper-parameter in \ourmodel:} We hereby specify the hyper-parameters when generalizing \ourmodel\ to new organs. The quantification process requires setting the typical \textit{highest and lowest HU values} for organs. The growth and interaction process requires setting the \textit{invasion pressure threshold} based on organ softness and the \textit{maximum growth steps} based on organ size. The mapping function requires setting the typical \textit{tumor HU values} for different organs.

\begin{table*}[t]
    \centering
    \scriptsize
    \caption{\textbf{Performance of liver, pancreatic, and kidney tumor segmentation.} \ourmodel\ outperforms Hu~\etal~\cite{hu2023label} and even surpass models trained on real tumors. Higher DSC and NSD values imply more precise overall segmentation; lower SD and HD values suggest more accurate segmentation boundaries.
    }
    \begin{tabular}{p{0.09\linewidth}p{0.12\linewidth}|P{0.12\linewidth}P{0.12\linewidth}|P{0.12\linewidth}P{0.12\linewidth}|P{0.12\linewidth}P{0.12\linewidth}}
        \toprule
         \multirow{3}{*}{organ}& \multirow{3}{*}{tumor} & \multicolumn{2}{c|}{U-Net} & \multicolumn{2}{c|}{Swin~UNETR} & \multicolumn{2}{c}{nnU-Net}   \\
         & &DSC/NSD &SD/HD &DSC/NSD &SD/HD &DSC/NSD &SD/HD\\
         &&(\%) $\uparrow$ & (mm) $\downarrow$ &(\%) $\uparrow$ & (mm) $\downarrow$ &(\%) $\uparrow$ & (mm) $\downarrow$ \\
        \midrule 
        \multirow{3}{*}{liver}  &  real tumors& 56.7/58.0 & 23.2/61.4 & 53.5/55.9 & 21.3/57.8 &56.2/55.3 & 24.6/58.3 \\ 
        
        & Hu~\etal & 54.5/57.6 & 23.8/58.8 & 52.3/56.5 & 22.9/56.9 & 53.7/56.1 & 22.5/57.2 \\
        
          & \ourmodel\ & \cellcolor{ired!20}58.9/63.7 & \cellcolor{ired!20}17.9/52.4 & \cellcolor{ired!20}56.7/62.5 & \cellcolor{ired!20}18.7/51.3 & \cellcolor{ired!20}57.9/63.2 & \cellcolor{ired!20}18.9/52.7 \\
          
          \midrule
          
        \multirow{3}{*}{pancreas}  &  real tumors & 57.8/56.5 & 13.1/47.7 & 56.7/52.8 & 24.6/53.9 &56.8/52.1& 14.5/44.6\\

        & Hu~\etal & 54.1/52.2 & 15.7/49.3 & 53.6/54.9 & 22.5/47.4 & 54.6/52.4& 17.1/48.0\\ 
        
          & \ourmodel\ & \cellcolor{ired!20}60.9/57.1 & \cellcolor{ired!20}12.4/43.5 & \cellcolor{ired!20}59.3/59.5 & \cellcolor{ired!20}20.4/40.7 &\cellcolor{ired!20}59.8/56.9 & \cellcolor{ired!20}13.3/41.4\\  \midrule
          
        \multirow{3}{*}{kidney}  &  real tumors & 71.3/62.8 & 27.2/64.3 & 70.7/61.2& 19.8/57.1 &65.2/58.1 & 25.6/59.3\\ 

        & Hu~\etal & 63.2/55.4 & 35.1/69.0 & 61.7/52.3 & 26.2/61.6 & 55.5/49.9 & 27.9/62.7\\

          & \ourmodel\  & \cellcolor{ired!20}73.2/65.0 &\cellcolor{ired!20}13.6/40.9  & \cellcolor{ired!20}73.9/63.5 & \cellcolor{ired!20}15.9/45.7 &\cellcolor{ired!20}67.6/60.1 & \cellcolor{ired!20}14.8/42.2\\ 
          
        \bottomrule
    \end{tabular}
    \begin{tablenotes}
        \item DSC - dice similarity coefficient; NSD - normalized surface dice.
        \item SD - surface distance; HD - Hausdorff distance. 
    \end{tablenotes}
    \label{tab:performance}
\end{table*}

\subsection{Results}

\smallskip\noindent\textbf{Clinical Validation:}
We performed a Visual Turing Test~\cite{geman2015visual} on 150 CT images, including 25 images with real tumors and 25 images with synthetic tumors in each organ test set.
The results in \tableautorefname~\ref{tab:reader_studies} show performance metrics for different radiologists. Radiologist R1 (3 years experience) only achieves specificity below $30\%$. R2 (7 years) exhibits specificity around $50\%$, confusing half of the synthetic tumors. Even R3 (10 years) misidentifies $47.1\%$ of synthetic tumors as real. This demonstrates the realistic tumor simulation achieved by \ourmodel.

\smallskip\noindent\textbf{Tumor Segmentation Performance:} We benchmark \ourmodel\ against the state-of-the-art modeling-based method~\cite{hu2023label} and real-tumor method. \tableautorefname~\ref{tab:performance} demonstrates \ourmodel's superiority in liver segmentation with DSC of $58.9\%$, and NSD of $63.7\%$. We also achieve superior performance with real tumors in the pancreas and kidney, with DSCs of $60.9\%$ and $73.2\%$, respectively.
Compared to models trained on real liver tumors, \ourmodel\ outperforms by $5.7\%$ in NSD. We also surpass Hu~\etal\ by $4.4\%$ in DSC and $6.1\%$ in NSD.

\smallskip\noindent\textbf{Superiority in Boundary Segmentation:}
\ourmodel\ generates synthetic tumors with absolutely precise tumor masks, while real data annotations often have inaccuracies at boundaries, leading to \textit{label noise} and boundary segmentation inaccuracy.
In \tableautorefname~\ref{tab:performance}, we apply distance metrics (NSD, SD, and HD). Our \ourmodel\ synthesis approach surpasses real liver tumors, achieving $6.1\%$ improvement in NSD and reductions of $24.7\%$ in SD and $10.8\%$ in HD. These results highlight the precision of \ourmodel\ in boundary segmentation, indicating its potential for surgical guidance, particularly in tumor excision procedures.

\begin{figure*}[t]
	\centering
\includegraphics[width=\linewidth]{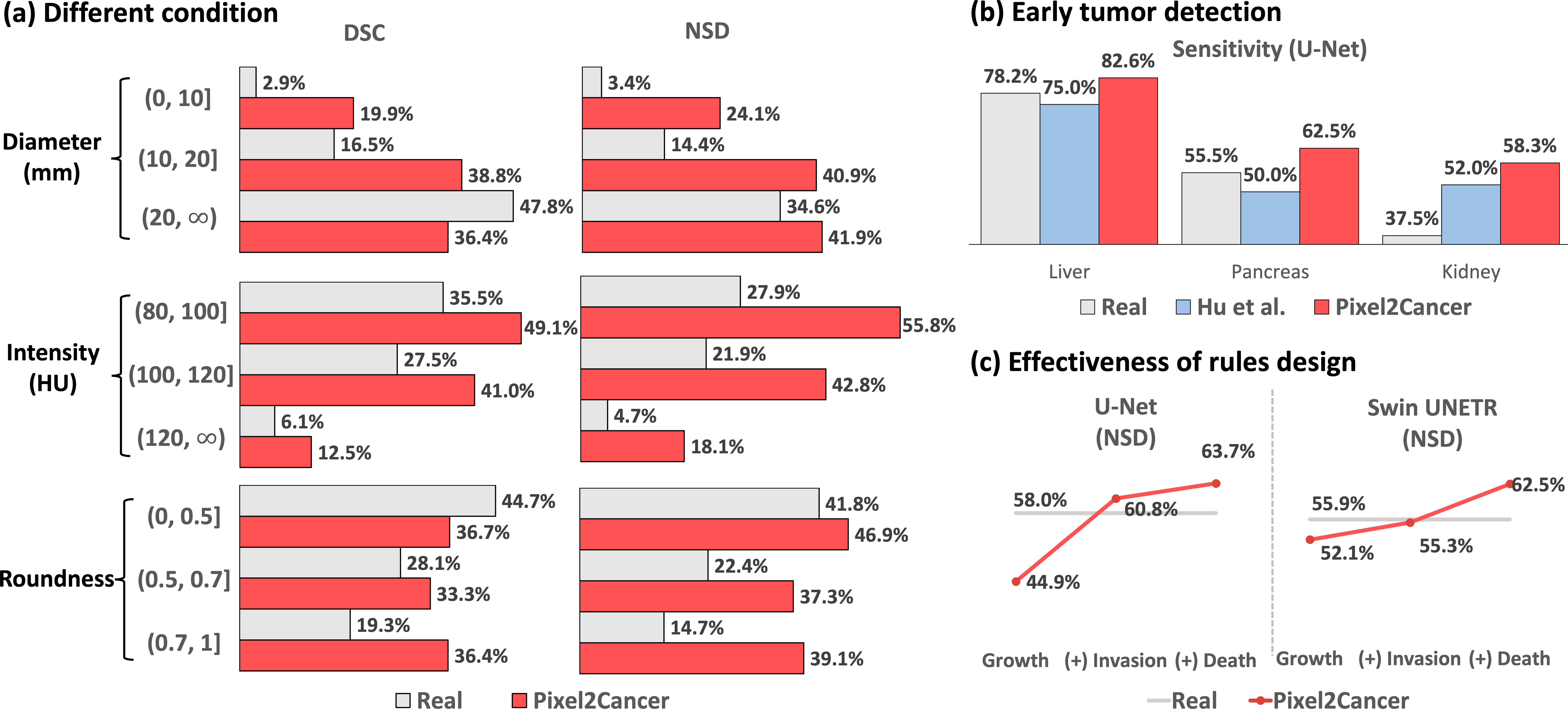}
\caption{\textbf{(a) Different condition:} We evaluate the influence of different tumor sizes, intensities, and roundness on liver tumors to assess the robustness of \ourmodel. \textbf{(b) Early tumor detection:} We report the Sensitivity of U-Net tumor detection among real tumors, Hu~\etal~\cite{hu2023label}, and \ourmodel\ in three organs. \textbf{(c) Effectiveness of rules:} We evaluated the performance of each generic rule on liver tumors, demonstrating the effectiveness of our rule design in tumor simulation.}
 \label{fig:ablation}
\end{figure*}

\smallskip\noindent\textbf{Ability in Small Tumor Detection:} Early detection of small tumors is essential for timely cancer diagnosis, but real datasets often lack sufficient instances due to the lack of CT scans in the early stages. \figureautorefname~\ref{fig:ablation}(b) and Appendix~\figureautorefname~\ref{fig:early_tumors} illustrates the performance of small tumor detection. Training solely on our synthetic tumor outperforms models trained on real tumors and Hu~\etal~\cite{hu2023label}, highlighting the potential of synthetic tumors for early cancer detection.

\smallskip\noindent\textbf{Ablation Studies:} (1) We evaluated the impact of various tumor conditions on the performance of the model. \ourmodel\ exhibits greater variability and robustness across different conditions compared to real tumors (\figureautorefname~\ref{fig:ablation}(a)). (2) We evaluated the effectiveness of generic rules on liver tumors (\figureautorefname~\ref{fig:ablation} (c)) Basic growth rules yielded $52.1\%$ NSD. Incorporating interaction and invasion matched real data performance. Adding the tumor death rule outperformed real tumors by $4.9\%$ in NSD, demonstrating the effectiveness of these generic rules.

\section{Conclusion \& Discussion}\label{sec:conclusion}

In this paper, we highlight the potential of \ourmodel\ for tumor synthesis, data augmentation, small tumor detection, and accurate boundary segmentation.
Despite simulating tumor behaviors and interactions with organ tissues, including the bending of vessels and the deformation of organ boundaries, there are still various mechanisms of accompanying changes in organs caused by tumors that we have not yet considered, including focal pancreas parenchymal atrophy
hepatic capsular retraction~\cite{tan2016causes}, and splenomegaly. In future work, we plan to devise organ rules within our approach to accurately simulate changes induced by tumors when generating synthetic tumors.

\begin{credits}
\subsubsection{\ackname} This work was supported by the Lustgarten Foundation for Pancreatic Cancer Research and the Patrick J. McGovern Foundation Award.

\subsubsection{\discintname}
The authors have no competing interests to declare that are relevant to the content of this article.
\end{credits}

%
%
%

\bibliographystyle{splncs04}
\bibliography{Paper-1596}
\appendix
\newpage

\begin{table*}[t]
    \centering
    \scriptsize
    \caption{\textbf{Early tumor performance.} In this evaluation, tumors are generally small or medium-sized, presenting predominantly challenging cases for tumor segmentation. This challenge is particularly pronounced in the pancreas and kidney, where tumors are more difficult to segment compared to liver tumors.}
    \begin{tabular}{p{0.09\linewidth}P{0.12\linewidth}P{0.16\linewidth}|P{0.142\linewidth}P{0.142\linewidth}|P{0.142\linewidth}P{0.142\linewidth}}
        \toprule
         \multirow{3}{*}{organ}& \multirow{3}{*}{tumor} & \multirow{3}{*}{label/non-label} & \multicolumn{2}{c|}{U-Net} & \multicolumn{2}{c}{Swin~UNETR} \\
        & & &DSC / NSD&SD / HD&DSC / NSD&SD / HD\\
        & & &(\%) $\uparrow$ &(mm) $\downarrow$&(\%) $\uparrow$&(mm) $\downarrow$\\
        \midrule
        \multirow{3}{*}{liver}  &  real tumors &101/0 &46.7/48.0 & 23.2/61.4 &  43.5/45.9 & 21.3/57.8\\ 
        
          & Hu~\etal~\cite{hu2023label} &0/116 & 44.5/47.6 & 23.8/58.8 & 42.3/46.5 & 22.9/55.4\\ 
          
          & \ourmodel\ &0/116 &  \cellcolor{ired!20}\textbf{47.2}/\cellcolor{ired!20}\textbf{52.9} & \cellcolor{ired!20}\textbf{17.9}/\cellcolor{ired!20}\textbf{52.4} & \cellcolor{ired!20}\textbf{45.4}/\cellcolor{ired!20}\textbf{51.6} & \cellcolor{ired!20}\textbf{18.7}/\cellcolor{ired!20}\textbf{51.3}\\ 
          
          \midrule
          
        \multirow{3}{*}{pancreas}  &  real tumors & 96/0&34.3/33.9 & 22.2/\textbf{47.7} & 28.3/33.6 & 22.3/45.9\\ 
        
          & Hu~\etal~\cite{hu2023label}  &0/104& 27.3/28.0 & 28.4/52.0 & 25.8/27.3 & 29.5/50.9 \\ 
          
          & \ourmodel\ &0/104 &\cellcolor{ired!20}\textbf{36.5}/\cellcolor{ired!20}\textbf{35.9} & \cellcolor{ired!20}\textbf{21.9}/\cellcolor{ired!20}52.2 & \cellcolor{ired!20}\textbf{31.5}/\cellcolor{ired!20}\textbf{34.6}& \cellcolor{ired!20}\textbf{21.7}/\cellcolor{ired!20}\textbf{41.8}\\ 
          
          \midrule
          
          \multirow{3}{*}{kidney}  &  real tumors &96/0 & 17.9/16.1 & 102.8/150.6 & \textbf{29.4}/27.6& 93.8/136.8\\ 
          
          & Hu~\etal~\cite{hu2023label} &0/120 & 14.8/17.4 & \textbf{70.1}/\textbf{112.9} & 23.5/25.9 & 83.2/\textbf{122.3} \\ 
          
          & \ourmodel\ &0/120 & \cellcolor{ired!20}\textbf{18.1}/\cellcolor{ired!20}\textbf{19.3} & \cellcolor{ired!20}85.2/\cellcolor{ired!20}131.7 & \cellcolor{ired!20}28.6/\cellcolor{ired!20}\textbf{28.2} & \cellcolor{ired!20}\textbf{80.9}/\cellcolor{ired!20}127.2\\
          
        \bottomrule
    \end{tabular}
    \label{tab:comparation_real}
\end{table*}

\begin{figure}[h]
  \centering
  \resizebox{\textwidth}{!}{\includegraphics[]{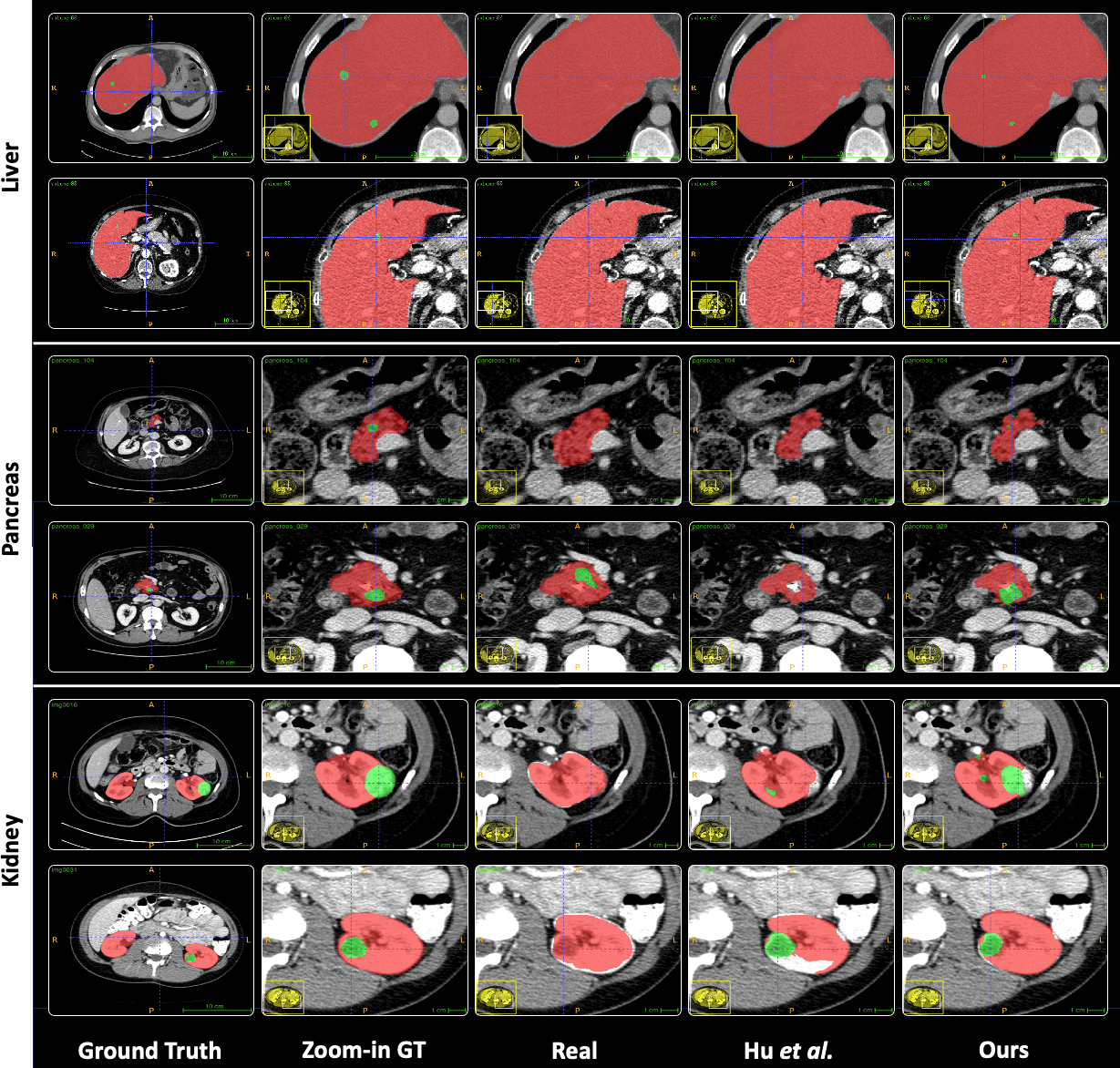}}
  \caption{\textbf{Examples of early tumor detection.} Qualitative visualizations of segmentation models for liver, pancreas, and kidney tumor detection. 
} \label{fig:early_tumors}
\end{figure}

\begin{figure}
\centering  
\resizebox{0.95\textwidth}{!}{\includegraphics[]{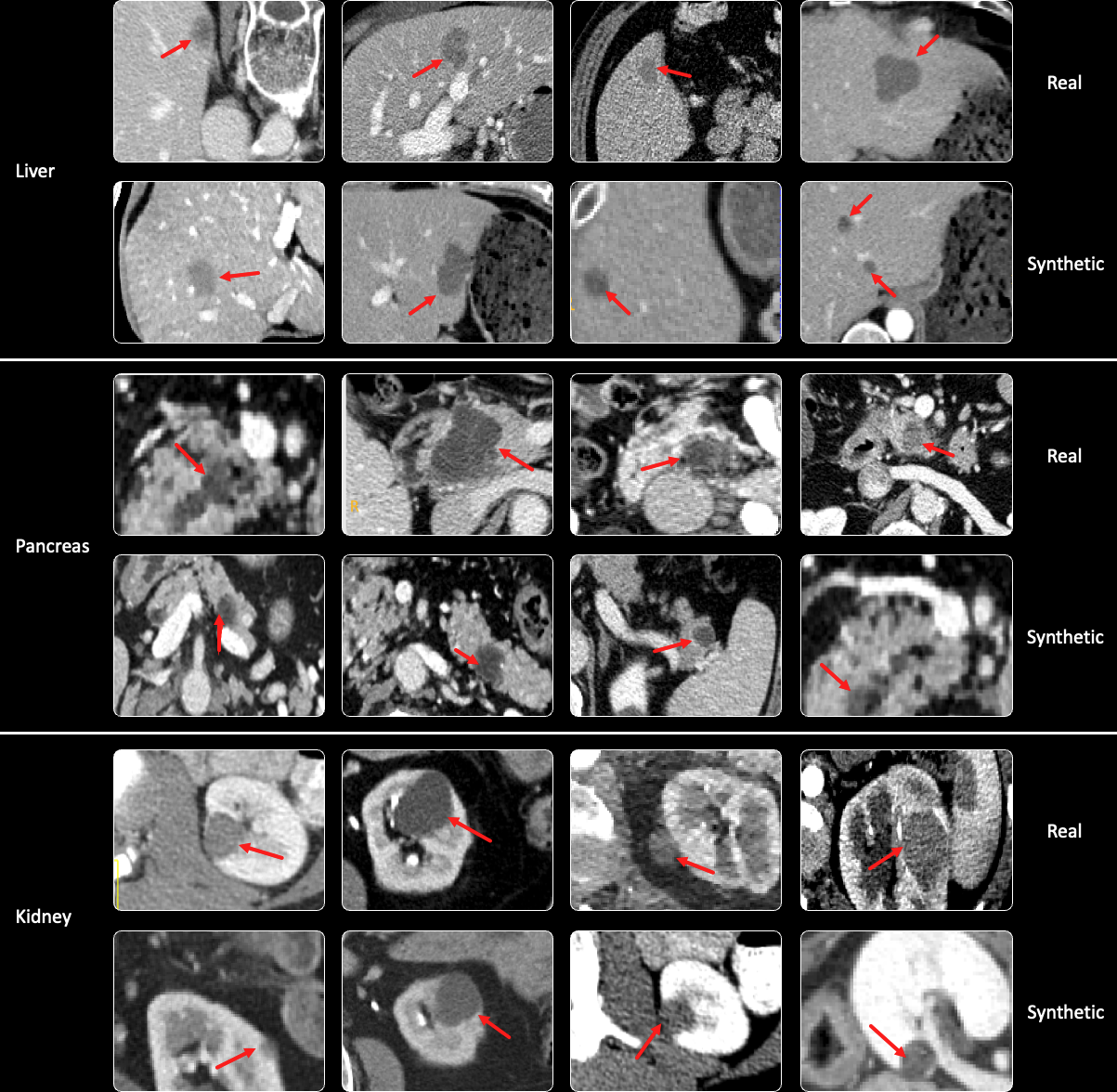}}

\caption{\textbf{Examples of Visual Turning Test.} Tubular results are presented in \tableautorefname~\ref{tab:reader_studies}. Our \ourmodel\ can be used to augment available healthy CT volumes.
} 
\label{fig_visal_turning_test}
\end{figure}

\begin{figure}
  \centering
  \resizebox{0.95\textwidth}{!}{\includegraphics[]{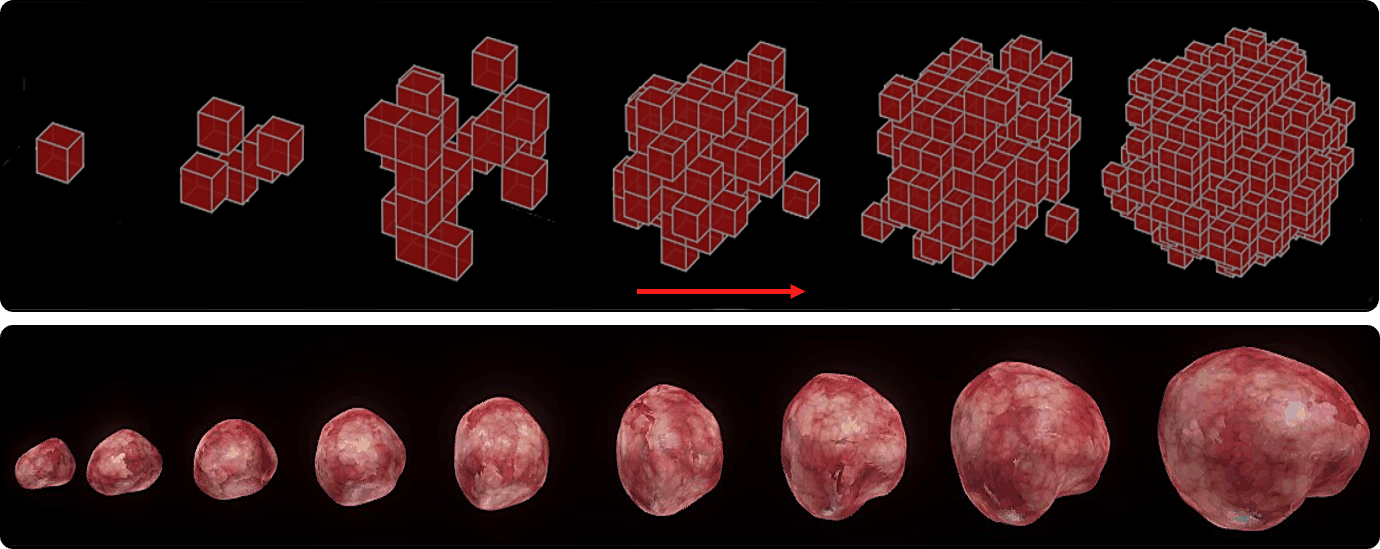}}
  \caption{\textbf{Tumor development visualization.} This rendering visualization depicts the iteration of \ourmodel\ and the progression of tumors. We illustrate iterations and the gradual development of tumors.
} \label{fig_development_visualization}
\end{figure}

\end{document}